\documentclass[10pt,letterpaper]{revtex4-1}
\usepackage{amsmath}
\usepackage{amssymb}
\usepackage{tabularx}
\usepackage{tabu}
\usepackage{graphicx}
\usepackage[toc]{appendix}
\usepackage{multirow}
\usepackage{comment}
\usepackage{makecell}
\usepackage{placeins}
\usepackage{subcaption}
\usepackage[export]{adjustbox}
\usepackage{xcolor}

\begin{document}

\title{Temperature Dependence of Beam on Plasma Stopping Power in the Resonance Regions of Fusion Reactions}
\author{Keh-Fei Liu} 
\affiliation{\mbox{Department of Physics and Astronomy, University of Kentucky, Lexington, KY 40506, USA}}

\begin{abstract}
A recent proposal of accelerator based fusion reactor considers a scheme where 
an ion beam from the accelerator hits the target plasma on the resonance of the fusion reaction so that
the reactivity ($\sigma v$) can be an order of magnitude larger than that of a thermonuclear reactor. 
One of the important inputs is the stopping power which is needed to assess the energy loss of the beam in the plasma.  In this work, we shall use
the analytic formulation of Brown, Preston and Singleton~\cite{Brown:2005ji} to calculate the temperature dependence of the stopping power
due to the target $t, {}^3H_e$, and ${}^{11}B$ plasmas in the 
resonance regions of their respective fusion reactions, i.e., $ d + t \rightarrow n + \alpha,
d + {}^3H_e \rightarrow p + \alpha$, and $p + {}^{11}B \rightarrow 3 \alpha$. It is found that
the calculated stopping power, especially when the quantum corrections are included, does not go down with temperature
as fast at $T^{-3/2}$. Instead it decreases slower, more like $T^{-x}$ with $x \le 1$
in the range of T from $\sim$ 5 to 50 keV for $d$ on $t$ and ${}^3H_e$ plasmas around their resonance energies.
\end{abstract}

\maketitle

\section{Introduction}

 Research to harvest energy from controlled fusion reaction has been a challenge for more 
than six decades. The thermonuclear reactor such as at ITER~\cite{ITER} is at a temperature (12.5 keV) which is much lower than that of the  peak resonance energy of the $d+t$ reaction with a center of mass energy of 64 keV. Thus, it is the exponential
tail of the Maxwell-Boltzmann distribution that is important in the integrated reaction rate $\langle \sigma v\rangle
\sim 10^{-22}\,{\rm m^3/s}$. Whereas,  direct $d$ on $t$ on the resonance yields a $\langle \sigma v\rangle
= 1.6 \times 10^{-21} m^3/s$ which is an order of magnitude larger.  In fact, many of the light ion fusion reactions 
have resonances at center-of-mass energy of 64 -- 300 keV with widths of 200 -- 400 keV. This has prompted the
proposal to explore the possibility of a fusion reactor with the fusion nuclei 
colliding at the energy where the fusion cross section peaks in order to maximize the reaction rate~\cite{Liu:2017pww}. This requires a beam at a particular energy. 

   However, the reaction rate (i.e. reactivity) is not the only concern for a reactor to work. All the possible energy losses
need to be taken into account. In particular, it is desirable to minimize the energy loss of the beam, which is measured in terms of the stopping power. It is pointed out that plasma has a special property in that the stopping power of a charged beam on plasma decrease as $T^{-3/2}$~\cite{Chu72,Hamada78}, where $T$ is the plasma temperature. As such, there should be a critical temperature, beyond which the power of fusion production will overcome the stopping power loss~\cite{Liu:2017pww}. This criterion for this  is characterized by the ratio
\begin{equation}   \label{R-sp}
R_{\rm sp} = \frac{\overline{\sigma Q} \epsilon_{\rm out}}{Z\,|dE/dx|(T)/(n_e\epsilon_b)} 
=  \frac{\overline{\sigma Q} n_e \, v_b\, \epsilon_{\rm out}}{Z|dE/dt|(T)/\epsilon_b} \ge 1.
\end{equation}
where $|dE/dx| (T)$ is the stopping power which depends on $T$.  $Q$ is the fusion energy gain and $\sigma$ is the fusion reaction cross section. $n_e$ is the electron density. $\epsilon_{out}$ is the output energy conversion efficiency to electricity.  $v_b$ is the beam velocity. For charged particle production, the direct conversion is possible which gives $\epsilon_{out} \sim 0.9$. For neutron production, 
$\epsilon_{out} \sim 0.3-0.4$.  $\epsilon_b$ is the efficiency to produce the beam at the resonance energy. 

   We have used the analytic calculation of the stopping power by L.S. Brown, D.L.Preston and R.L.Singleton, Jr. (BPS)~\cite{Brown:2005ji} at certain temperature and used the $T^{-3/2}$ rule to determine the critical temperature
$T_c$ for Eq.~(\ref{R-sp}) for the $ d + t \rightarrow n + \alpha,
d + {}^3H_e \rightarrow p + \alpha$, and $p + {}^{11}B \rightarrow 3 \alpha$ reactions in a previous publication~\cite{Liu:2017pww}
and obtained fairly low $T_c$ as compared to those needed in the thermonuclear reactors. 
In view of the fact that the stopping power is an important
quantity which will determine the heating rate of the plasma, the length of the reactor, the plasma lifetime and, the critical
flux of the incoming beam to achieve a net energy gain, it is a critical parameter in the practical design of the accelerator based fusion reactor (ABFR). Not many experiments are performed for the case where $v_b$ is close to the electron thermal velocity $v_e^{th}$ in the plasma. One recent effort~\cite{Cayzac2017} where the experimental set up
includes a nitrogen projectile at an energy of $E_p =0.586 \pm 0.016$ MeV per nucleon, the plasma electron density
$n_e \simeq 5 \times 10^{20}\, {\rm cm^{-3}}$, and electron temperature $T_e \simeq 150$ eV (so that
$v_b/v^e_{th} \simeq 1.2$) was carried out to measure the stopping power near the Bragg peak. It was found that
the results agree well with the BPS prediction. Another experimental stopping power measurement was carried out for the 
low-Z ions at electron temperature and number densities in the range of 1.4 - 2.8 keV and 4 - 8 $\times 10^{23} {\rm cm^{-3}}$~\cite{Frenje2019}. Both low and high ion speeds were considered. Again, the BPS prediction provided a better
description of the ion stopping than other formalisms, except at $v_i \sim 0.3\, v^{th}$ where BPS formalism underpredicts by $\sim 20\%$. 

In this work, we shall directly calculate the analytic expression of BPS numerically and will include the ions in addition to the electrons in the stopping power. We shall study the temperature dependence of the ion stopping powers of $d$ on $t$, ${}^3 H_e$ and $d$ plasmas and $p$ on ${}^{11}B$ plasma. They are the favorable fusion reactions channels and the stopping powers of the ion beams on the respective plasma are critical parameters for the design
of ABFR.

\section{BPS Formalism of the Stopping Power}

    The BPS approach is an exact formalism to calculate the stopping power of a charged particle from the Fokker-Plank equation to the leading order of  the dimensionless plasma coupling parameter
$g= e^2 \kappa_D/4\pi T$ where $\kappa_D$ is the Debye wave number of the plasma.  The usual method to calculate
the stopping power of a charged particle through matter consists of two parts: the long-distance, soft collisions part is evaluated
from the $\vec{j}\cdot \vec{E}$ energy loss in a dielectric medium and the hard collisions are calculated through the Coulomb
scattering. Both involve logarithms and the sum does not give the coefficient inside the large logarithm, which is hard to
compute. A quantum field theory approach in terms of  dimensional continuation has been introduced to address this problem~\cite{Brown:2005ji}. The Lenard-Balescu kinetic equation describes the long-distance, collective
behavior of the plasma, its stopping power does not encounter ultraviolet divergence for the spatial dimension $ \nu < 3$.
On the other hand, the energy loss from the short-distance collision of the Coulomb scattering over all impact parameters
is free of infrared divergence for $\nu > 3$. Both of them are divergent at the physical dimension $\nu =3$ with a simple pole and can be analytically continued beyond their original region of validity to the $\nu > 3$ and $\nu <3$ regions as a subleading contribution to leading order of the perturbation. Thus, the sum of the two gives the leading and subleading contributions for all $\nu$. At $\nu = 3$, the pole terms cancel and give a finite logarithm term with the correct coefficient inside the log term. It has the form of $A g^2 \ln B g^2 + \mathcal{O}(g^3)$ with the correct coefficients of both $A$ and $B$.

     We shall use the BPS results to study the temperature dependence of the energy loss of the beam due to both
the electrons and ions in the plasmas for $d$ on $t$, ${}^3 H_e$ and $d$ plasmas and $p$ on ${}^{11}B$ plasma relevant to the reactions $ d + t \rightarrow n + \alpha, d + {}^3H_e \rightarrow p + \alpha$, and $p + {}^{11}B \rightarrow 3 \alpha$ with the beam energies at their respective resonance energies. Even though the bulk motion of the plasma may not be small relative to the that of the bombarding charged particle,
we shall consider the situation with a stationary plasma as in the case of the BPS formalism. The stopping power in the BPS formalism has been divided into several contributions.


\begin{eqnarray}  \label{classical}
\frac{dE_b^C}{dx} = \frac{dE_{b,S}^C}{dx}+\frac{dE_{b,R}^{<}}{dx}, \\
\end{eqnarray}
where $\frac{dE_b^C}{dx}$, the classical contribution, is made of  the sum of $\frac{dE_{b,S}^C}{dx}$ which
is the Coulomb scattering contribution regulated to have the singular pole piece cancelled from dimensional
continuation and $\frac{dE_{b,R}^{<}}{dx}$ which is from the regular contribution from $\nu < 3$. 
There is a quantum correction. Thus, for species b, the total energy loss is
\begin{equation}
\frac{dE_b}{dx} = \frac{dE_{b}^C}{dx}+\frac{dE_{b}^{Q}}{dx}. 
\end{equation}
where $\frac{dE_{b}^{Q}}{dx}$ is the quantum scattering contribution in the Born approximation for the
species b.
The total energy loss is the sum due to all species in the plasma.
\begin{equation}
\frac{dE}{dx}=\sum_{i}{\frac{dE_{b_i}}{dx}}
\end{equation}

Before we give the expressions of each of the above energy loss terms, we shall list the variables used
in the formulas and their descriptions. They are listed in Table~\ref{1}.

\begin{table}[ht] \label{1}
\centering
\includegraphics[width=4.5in, height=3in]{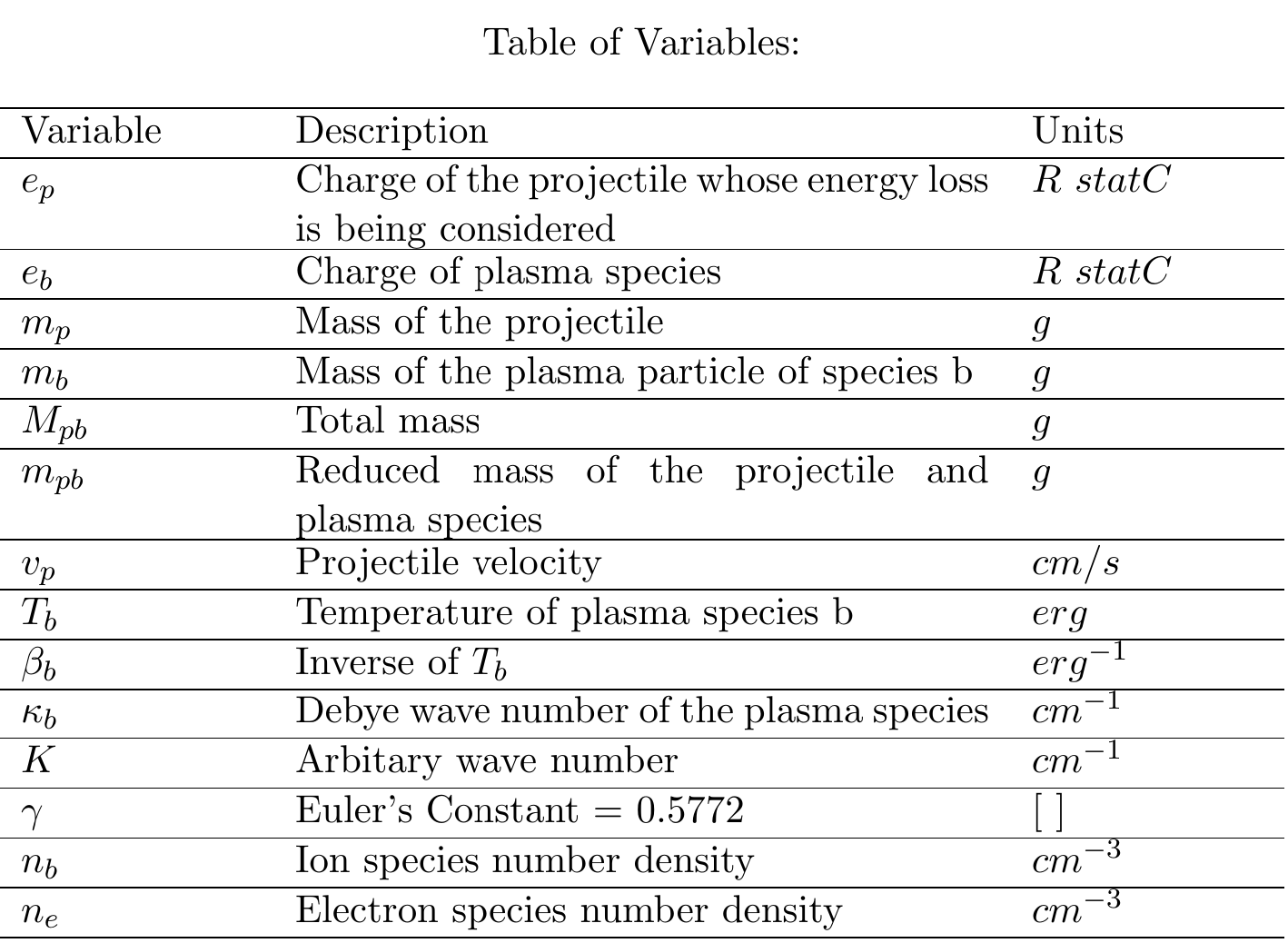}
\caption{Table of variables used in the formulas for the energy loss.}
\label{1}
\end{table}

The Coulomb scattering part $\frac{dE_{b,S}^C}{dx}$ in Eq.~\ref{classical}  is 

$$ \frac {dE^{C}_{b,S}}{dx}=\frac{e_p^2}{4\pi}\frac{\kappa _b^2}{m_p v_p}\left( \frac{m_b}{2\pi\beta_b}  \right)^{1/2} \int _{0}^{1} du u^{1/2} exp \left[ - \frac{1}{2} \beta_b m_b v_p^2 u \right]$$
$$ \text{ x }\left( \left[ -ln\left( \beta_b \frac{e_p e_b K}{4\pi}\frac{m_b}{m_{pb}}\frac{u}{1-u} \right) +2-2\gamma
\right] \left[ \beta_b M_{pb} v_p^2 -\frac{1}{u} \right]+\frac{2}{u} \right)  $$\\

and $\frac{dE_{b,R}^{<}}{dx}$ in Eq.~\ref{classical} is

$$ \frac{dE^{<}_{b,R}}{dx}=\frac{e_p^2}{4\pi}\frac{i}{2\pi}\int_{-1}^{+1}dcos\theta\, cos\theta\frac{\rho_b\left( v_p cos\theta \right)}{\rho_{total} \left( v_p cos\theta \right)} F\left( v_p cos\theta \right) ln\left( \frac{F\left( v_p cos\theta \right)}{K^2} \right) -\beta $$

where

\begin{eqnarray}
F(u) &=&-\int_{-\infty}^{+\infty}dv\frac{\rho_{total}(v)}{u-v+i\eta}  \\
 \rho_{total}(v) &=&\sum_{c}\kappa_c^2 v \sqrt{\frac{\beta_c m_c}{2\pi}} exp\left( -\frac{1}{2}\beta_c m_c v^2 \right)  \\
 \beta &=& \frac{e_p^2}{4\pi}\frac{i}{2\pi}\frac{1}{\beta_b m_p v_p^2}\frac{\rho_b\left( v_p \right)}{\rho_{total}\left( v_p \right)}\left[ F(v_p) ln\left( \frac{F(v_p)}{K^2}-F^*(v_p) ln\left( \frac{F^*(v_p)}{K^2} \right) \right) \right] 
\end{eqnarray}












After having coded the equations in Matlab, we have checked against BPS's results shown in the figures~\cite{Brown:2005ji} in their paper and they are replicated. 
The ion and electron plasma species have separate stopping powers which are calculated and added together to give the total stopping power of the plasma on the incoming projectile. We only consider neutral plasmas since they are inherently easier to confine. 

\section{Results}

\begin{figure}[htb]
\centering
\includegraphics[width=2.5in, height=2.1in]{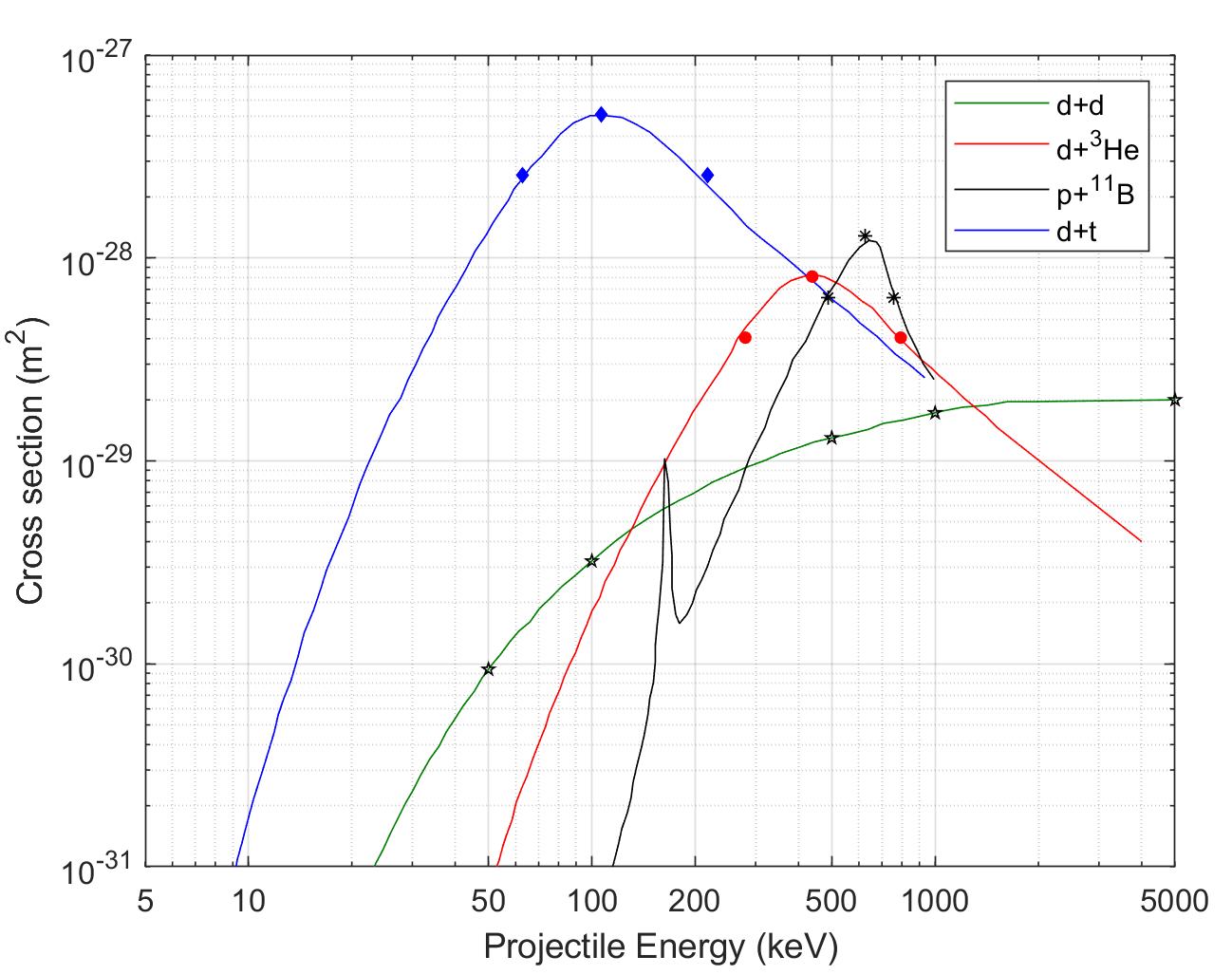}
\caption{Cross sections of the  $ d+d \rightarrow n+ {}^3H_e$, $d + t \rightarrow n + \alpha, d + {}^3H_e \rightarrow p + \alpha$, and $p + {}^{11}B \rightarrow 3 \alpha$ reactions as a function of the projectile laboratory energy.  We mark the
resonance energies and their respective $E_L$ and $E_H$ as explained in the text. We also calculate at 5 projectile energies for the $dd$ channel as described in the text.}
\label{fig:Xsec}
\end{figure}


ABFR utilities the large fusion cross sections in each reaction channel at their respective resonance energies. We show
the cross sections of the $ d+d \rightarrow n+ {}^3H_e$ and $p + t$, $d + t \rightarrow n + \alpha, d + {}^3H_e \rightarrow p + \alpha$, and $p + {}^{11}B \rightarrow 3 \alpha$ reactions as a function in the projectile laboratory energy in Fig.~\ref{fig:Xsec}. We see that there are fairly broad resonances in the $d+t$ and $d+{}^3H_e$ channels and relatively narrow resonance in the $p + {}^{11}B$ channel. There is no peak resonance structure in the $d+d$ channel.

     We shall present our calculations of the stopping powers for the energies at the resonances where the cross sections are at their maximums. We also calculate them at $E_{\rm L}$ and $E_{\rm H}$ which are the lower and higher boundaries of the full width at half maximum (FWHM) where the cross sections are at half of the maximum. These energies and the maximum cross sections 
 $\sigma_{\rm max}$ are tabulated in Table~\ref{tab:resonances}. These energies are marked in Fig.~\ref{fig:Xsec} where
 the cross sections for $ d+d \rightarrow n+ {}^3H_e$, $d + t \rightarrow n + \alpha, d + {}^3H_e \rightarrow p + \alpha$, and $p + {}^{11}B \rightarrow 3 \alpha$ reactions are shown as a function in the projectile laboratory energy. 
 For the $dd$ reaction, there is no resonance. We calculate the stopping power for the projectile energies at
 50, 100, and 500 keV, and also at 1 and 5 MeV.
 \begin{table}[ht]
\centering{}%
\begin{tabular}{cccccc}
Fusion Reaction Channel & $E_L$ (keV) & $E_{\rm R}$ (keV)& $E_H$ (keV)& $\Gamma$ (kev) & $\sigma_{\rm max}$(b) \tabularnewline
\hline 
{$d+t$} & 63 & 107 & 217 &  210 & 5.1\tabularnewline
\hline 
{$d + {}^3H_e$} & 280 &  439 & 794 & 430 &  0.81\tabularnewline
\hline
$ p + {}^{11} B$ &  487 & 625 & 756 & 300  & 1.28 \tabularnewline
\hline 
\end{tabular}
\caption{ The stopping powers for the three fusion channels are calculated at the resonance energies $E_{\rm R}$ in the lab frame and also at at $E_{\rm L}$ and $E_{\rm H}$ which are the lower and higher boundaries of the full width at half maximum (FWHM) where the cross sections are at half of the maximum. 
\label{tab:resonances}}
\end{table}


We first show the separate stopping powers due to the electrons and ions in the plasma as a function of the temperature at respective resonance energies as listed in Table~\ref{tab:resonances}. 

\begin{figure}[htbp]
\centering
\begin{subfigure}{0.4\textwidth}
{{\includegraphics[width=2.75in, height=2in]{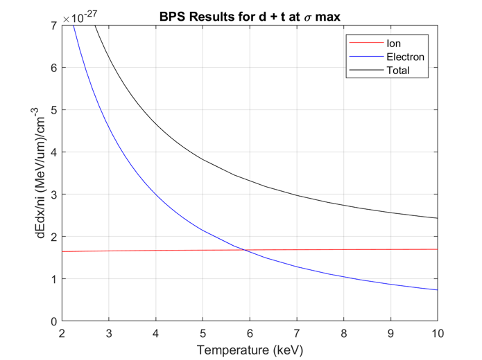}}
\caption{}
\label{fig:subim1}}
\end{subfigure}
\begin{subfigure}{0.4\textwidth}
{{\includegraphics[width=2.75in, height=2in]{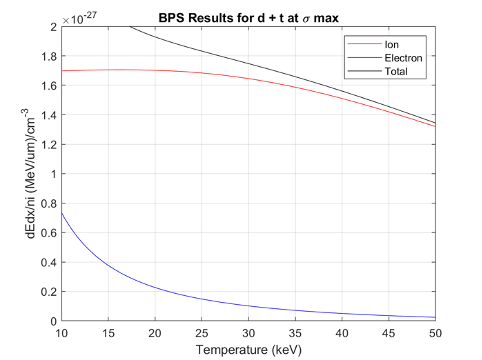}}
\caption{}
\label{fig:subim2}}
\end{subfigure}
\begin{subfigure}{0.4\textwidth}
\includegraphics[width=2.75in, height=2in]{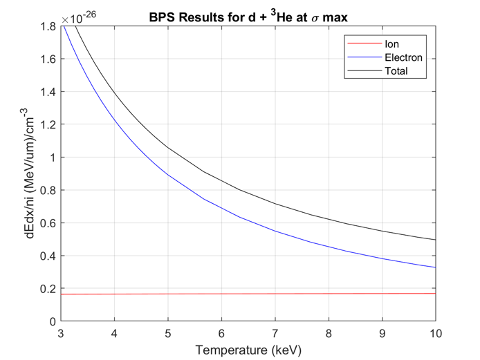} 
\caption{}
\label{fig:subim3}
\end{subfigure}
\begin{subfigure}{0.4\textwidth}
\includegraphics[width=2.75in, height=2in]{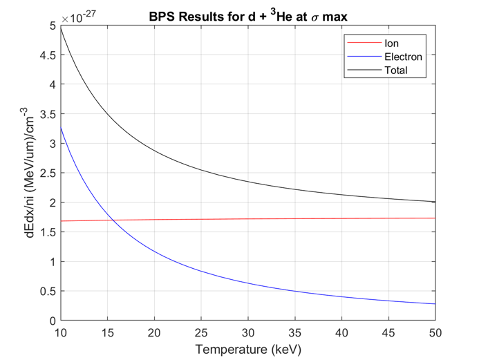}
\caption{}
\label{fig:subim4}
\end{subfigure}
\caption{The stopping powers due to the electrons and ions respectively as a function of temperature for
the $d$ on $t$ plasma at (a) $T = 2 - 10$ keV, (b) $T = 10 - 50$ keV, and $d $ on ${}^3H_e$ plasma at (c) $T = 2 - 10$ keV, (d) $T = 10 - 50$ keV.}
\label{ei-dt}
\end{figure}
We see in Fig.~\ref{ei-dt} for the $d$ on $t$ and ${}^3 H_e$ plasmas that below $T <  6$ keV, the stopping power due to the electrons in the $d$ on $t$ case is larger than that
due to the tritium ions and the ion stopping power dominate for $T \geq 20$ keV. The corresponding temperatures
for the $d$ on ${}^3H_e$  case are higher. The ion's contribution is catching up with that of the electrons at $\sim 15$ keV
and dominate after $\sim 40$ keV.

\begin{figure}[htbp]
\centering
\begin{subfigure}{0.4\textwidth}
\includegraphics[width=2.75in, height=2in]{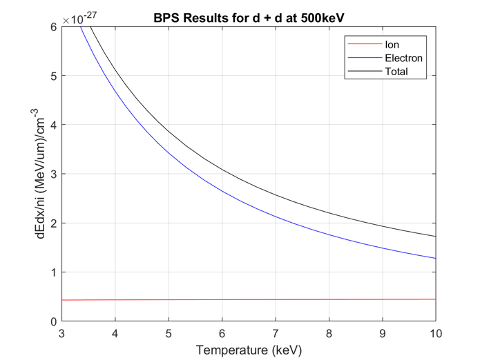} 
\caption{}
\label{fig:subim5}
\end{subfigure}
\begin{subfigure}{0.4\textwidth}
\includegraphics[width=2.75in, height=2in]{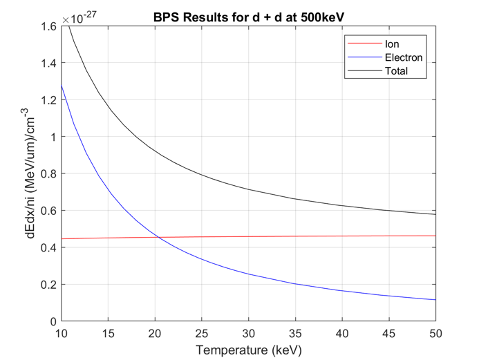}
\caption{}
\label{fig:subim6}
\end{subfigure}
\begin{subfigure}{0.4\textwidth}
\includegraphics[width=2.75in, height=2in]{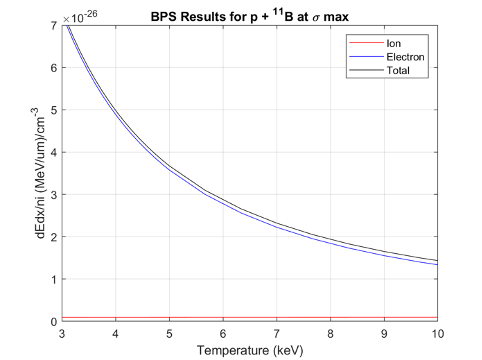} 
\caption{}
\label{fig:subim7}
\end{subfigure}
\begin{subfigure}{0.4\textwidth}
\includegraphics[width=2.75in, height=2in]{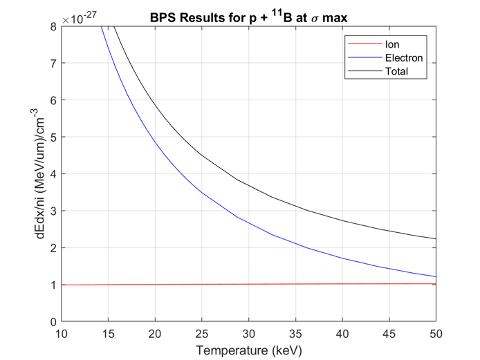}
\caption{}
\label{fig:subim8}
\end{subfigure}
\caption{The same as in Fig.~\ref{ei-dt} for the $d$ on $d$ plasma ((a) and (b)) with the $d$ projectile at 500 keV and $p$ on ${}^{11}B$ plasma ((c) and (d)) for the $p$ projectile at $E_R = 625$ keV.}
\label{ei-pB}
\end{figure}

Similar to Fig.~\ref{ei-dt}, we show in Fig.~\ref{ei-pB} $d$ on $d$ plasma results for the projectile $d$ at 500 keV and $p$ on ${}^{11}B$ plasma results  for the $p$ projectile at $E_R = 625$ keV.
While the relative electron and ion contributions of the $d$ on $d$ plasma in Fig.~\ref{ei-pB} are similar to those of $d$ on ${}^3H_e$ in Fig~\ref{ei-dt},
the ion stopping powers in $p$ on ${}^{11}B$ are much smaller than those of the electrons until $T > 50$ keV.

 \begin{figure}[htbp]
\centering
\begin{subfigure}{0.4\textwidth}
\includegraphics[width=2.75in, height=2in]{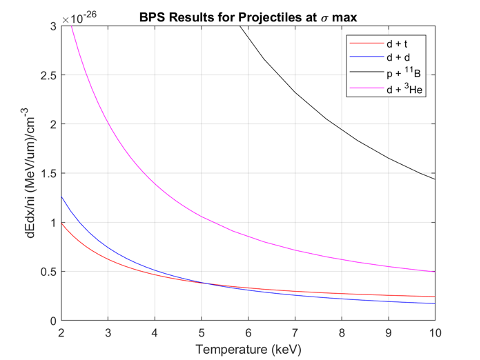} 
\caption{}
\label{fig:subim5}
\end{subfigure}
\begin{subfigure}{0.4\textwidth}
\includegraphics[width=2.75in, height=2in]{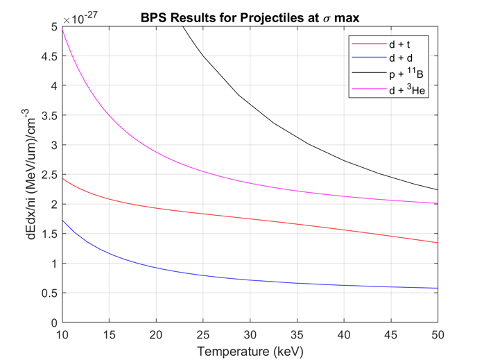}
\caption{}
\label{fig:subim6}
\end{subfigure} 
\caption{The total stopping powers from  both the electrons and ions as a function of temperature for
the  $d$ on $d, t, {}^3H_e$ plasmas and $p$ on ${}^{11}B$ plasma at (a) $T = 2 - 10$ keV, (b) $T = 10 - 50$ keV.}\label{compare}
\end{figure}
%
\newpage

  We compare the total electron and ion stopping powers in $d$ on $t, {}^3H_e$, and $d$ plasmas, and $p$ on ${}^{11}B$ plasma in Fig.~\ref{compare}. We see that $p$ on ${}^{11}B$ has the largest stopping power. The next is $d$ on ${}^3H_e$. $d$ on $t$ stopping power is close to that of $d$ on $d$ at 500 keV incoming energy for $T = 2 - 10$ keV and becomes larger for hight $T$.  \\
  


\begin{figure*}[htb]
\centering
\begin{subfigure}{0.47\textwidth}
\includegraphics[width=2.75in, height=2.25in]{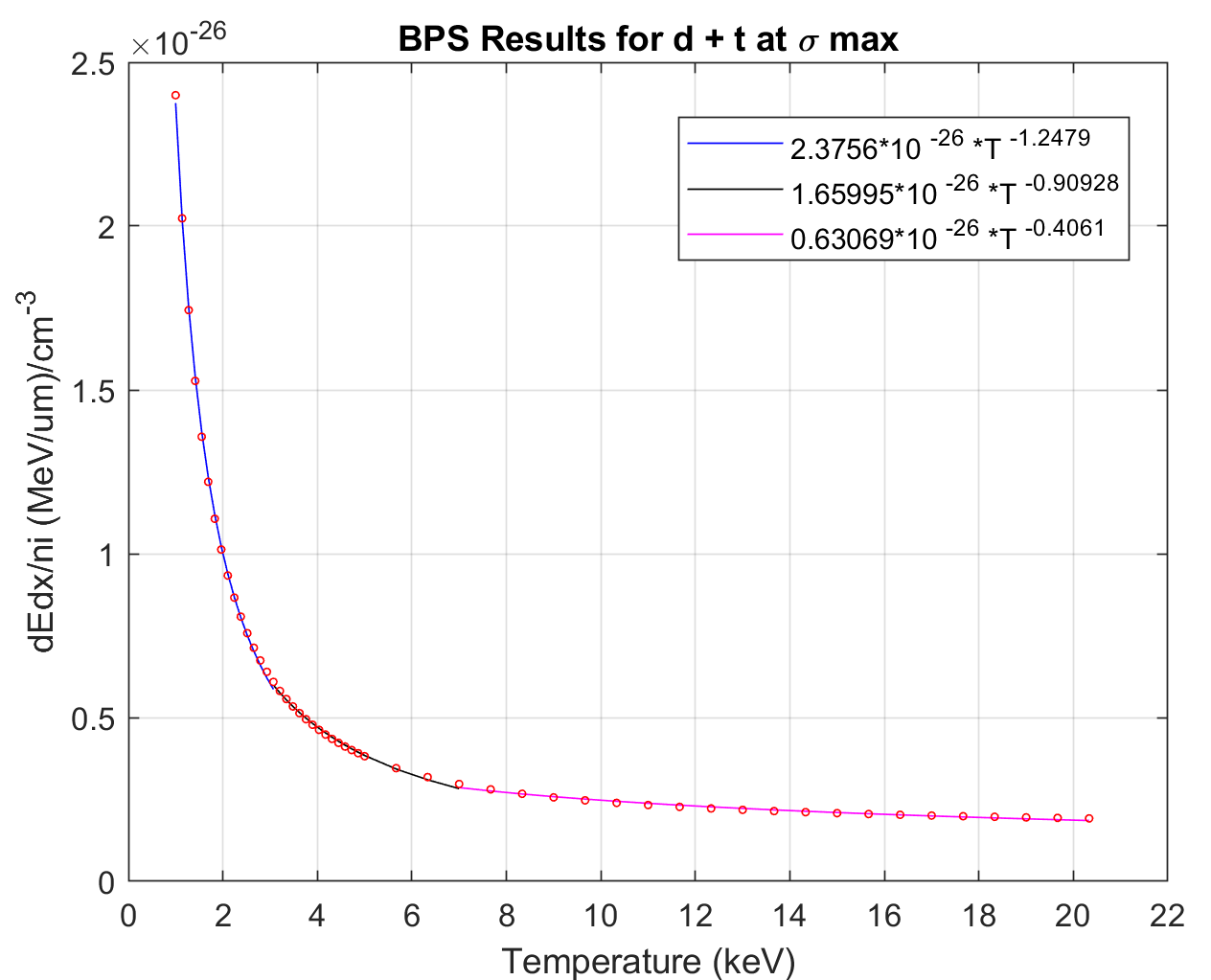} 
\caption{}
\label{tab:subim1}
\end{subfigure}
\begin{subfigure}{0.47\textwidth}
\includegraphics[width=2.75in, height=2.25in]{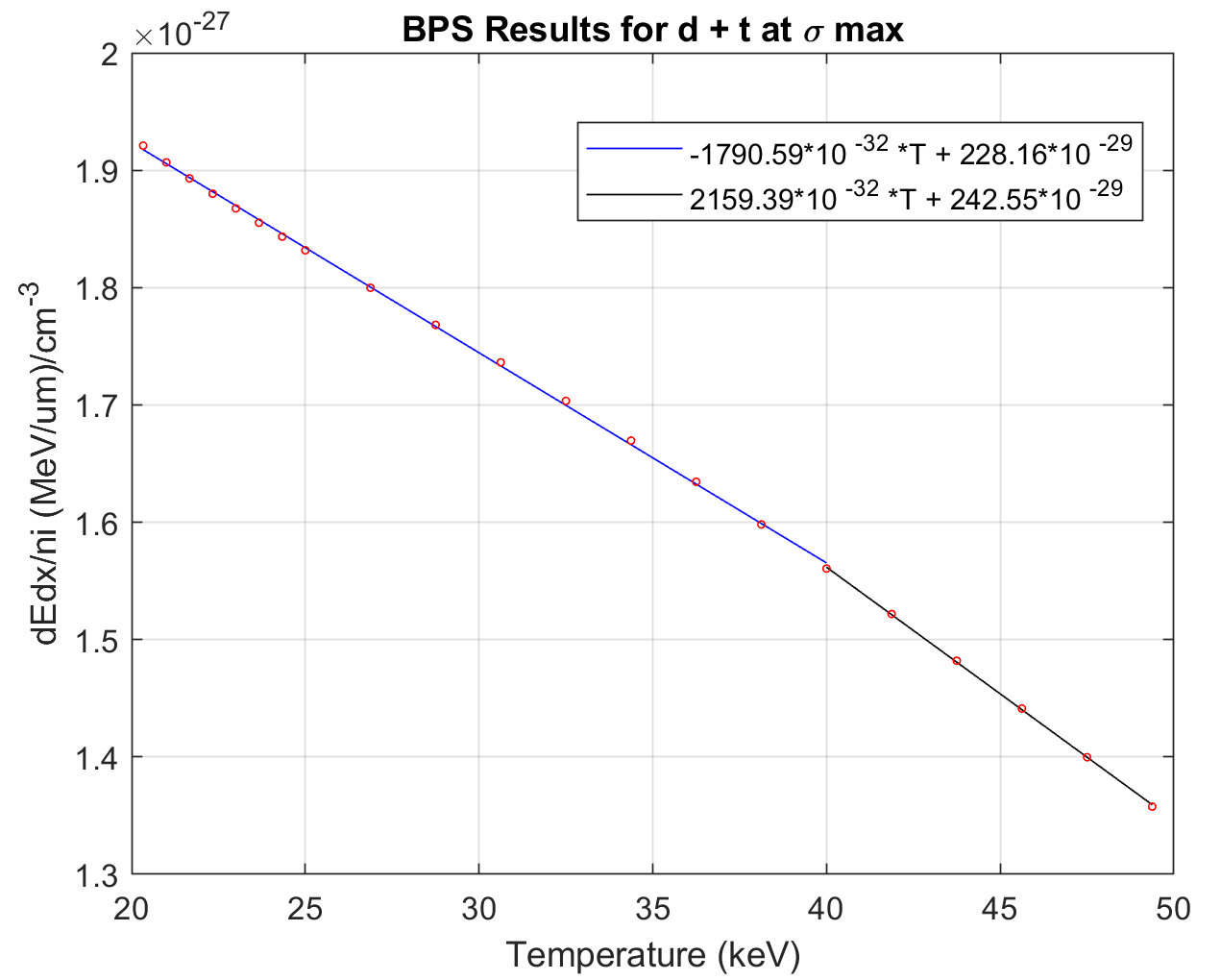} 
\caption{}
\label{tab:subim2}
\end{subfigure}
\caption{ (a)  Power-law fits of the data in the temperature range between 2 and 20 keV. (b) Linear regression fits of the data in the temperature range between 20 and 50 keV.}
\end{figure*}

\begin{table*}[htb]
\centering
\begin{subfigure}{0.47\textwidth}
\includegraphics[width=2.75in, height=4.25in]{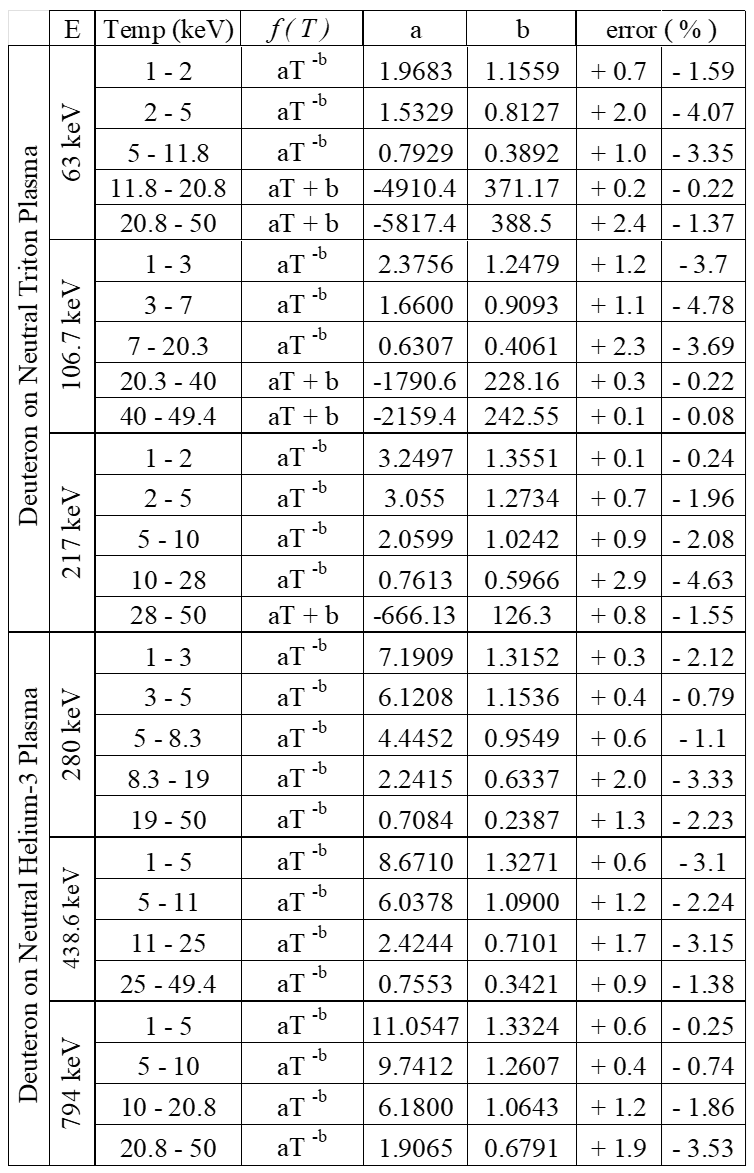} 
\caption{}
\label{tab:1}
\end{subfigure}
\begin{subfigure}{0.47\textwidth}
\includegraphics[width=2.75in, height=4.25in]{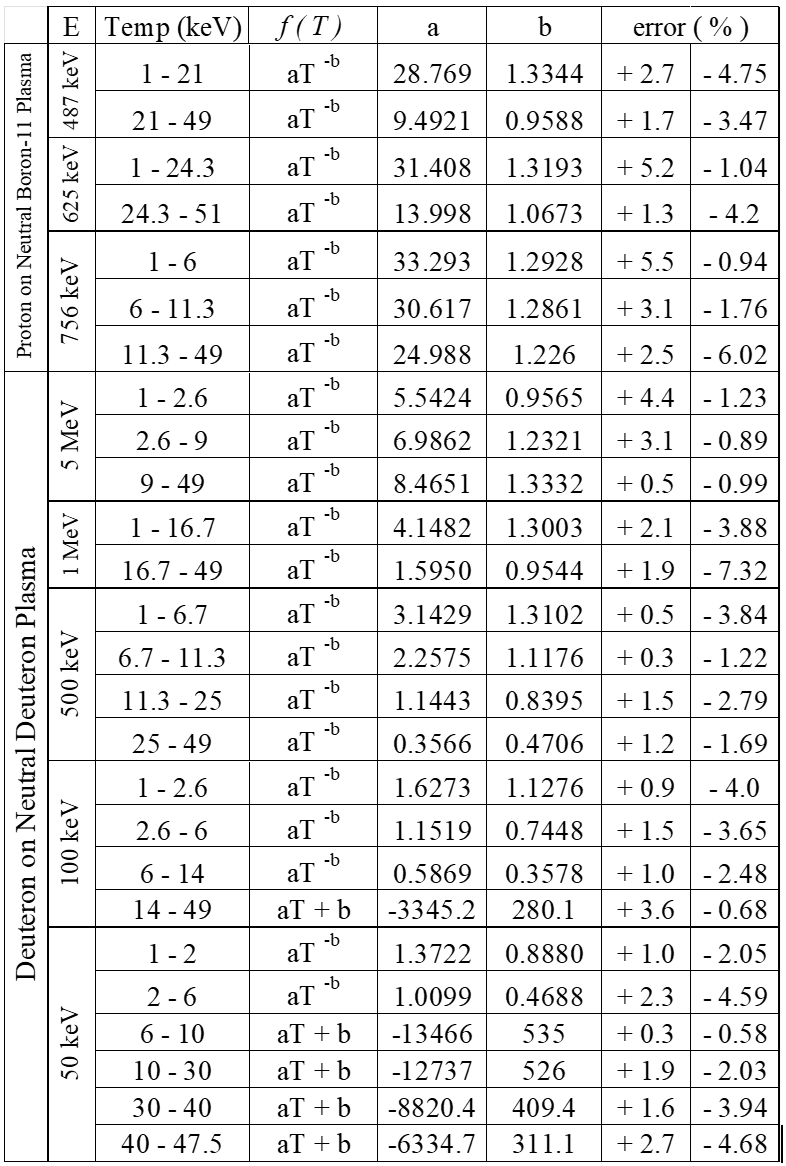}
\caption{}
\label{tab:2}
\end{subfigure}
\caption{The two tables above display the fitted results and the errors of the fits. Only the maximum and minimum errors are reported for the entire fitted temperature range. The stopping power 
$f(T)$ is in the unit of 
$ {\rm (MeV/\mu m)\, cm^{3}}$. $a$ is in the unit of $ 10^{-26} {\rm (MeV/\mu m)\, cm^{3} \, keV^b}$ for the power law and in the unit of $ 10^{-32} {\rm (MeV/\mu m)\, cm^{3}}$  for the linear regressions. $T$ is in keV. $b$ is a constant for the power laws, but for linear regressions it should be multiplied by $10^{-29} {\rm (MeV/\mu m)\, cm^3}$.   \label{Table-fit}}
\end{table*}


To facilitate future designs of ABFR, we parameterize them in analytic expressions in the form of $aT^{-b}$ and $aT + b$ for different region of $T$ with errors less than 5\%. The fit with the power form for $T < 20$ keV for the $d$ on $t$ plasma is given in Fig.~\ref{tab:subim1} and the linear fit for the range of $T$ between 20 keV and 50 keV is shown in Fig.~\ref{tab:subim2}. 
We tabulate them in Table~3 together with the errors. We see that in all the cases studied so far, the power $b$ in $aT^{-b}$ is less than $3/2$ as predicted~\cite{Chu72}. For the on resonance incoming energies, it is $\sim 1.3$ at low temperature, i.e., 1 - 3 keV for $d + t$, 1 - 5 keV for $d+{}^3H_e$ and 1 - 21 keV for $p + {}^{11}B$. It then decreases to around 1 and down to $\sim 0.4$ in 7 - 20 keV for $d + t$,
$\sim 0.3/1.1$ in 25 - 50 keV for $d + {}^3H_e/p + {}^{11}B$. We also parametrize the stopping powers at $E_L$ and $E_H$ (given in Table~\ref{tab:resonances} which are the lower and higher boundaries of the full width at half maximum (FWHM) where the cross sections are at half of the maximum to check for beam energy dependence. Same fits are done for the 5 energies in the $d$ on $d$ case.

\begin{figure}[htbp]
\centering
\begin{subfigure}{0.4\textwidth}
{\includegraphics[width=2.5in,height=1.7in]{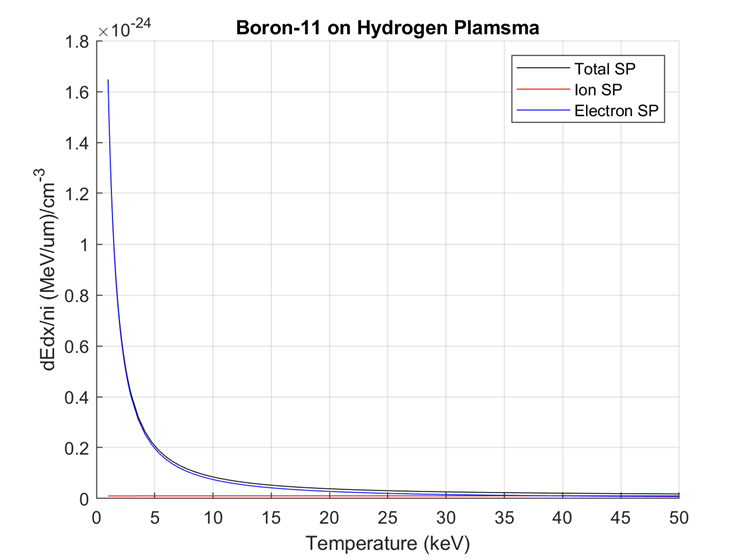}}
\caption{}
\label{fig:B11_p}
\end{subfigure}
\begin{subfigure}{0.4\textwidth}
{{\includegraphics[width=2.5in, height=1.7in]{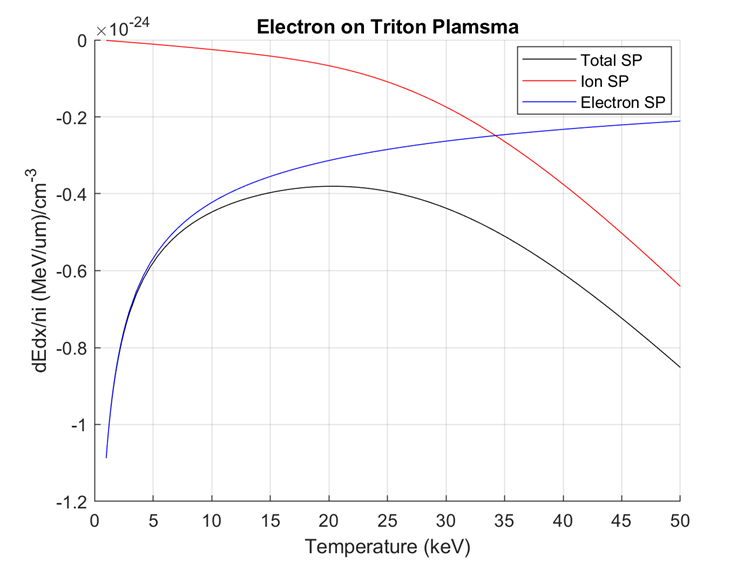}}
\caption{}
\label{e-T}}
\end{subfigure}
\begin{subfigure}{0.4\textwidth}
\includegraphics[width=2.5in, height=1.7in]{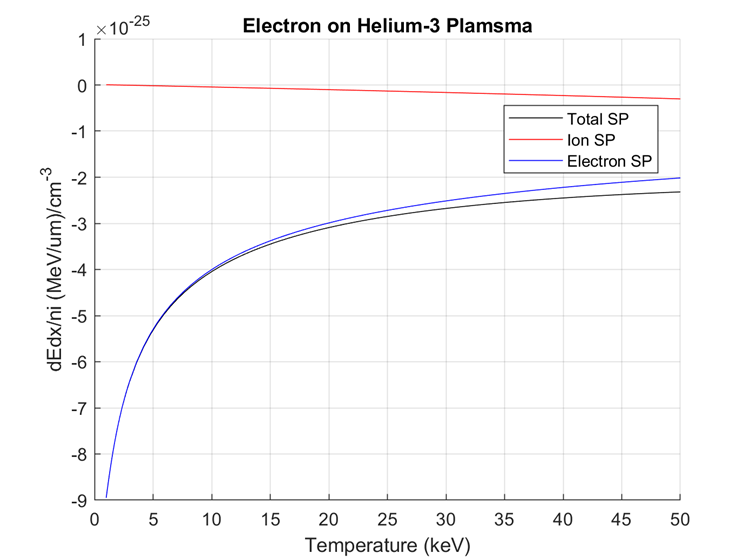} 
\caption{}
\label{e-He3}
\end{subfigure}
\begin{subfigure}{0.4\textwidth}
\includegraphics[width=2.5in, height=1.7in]{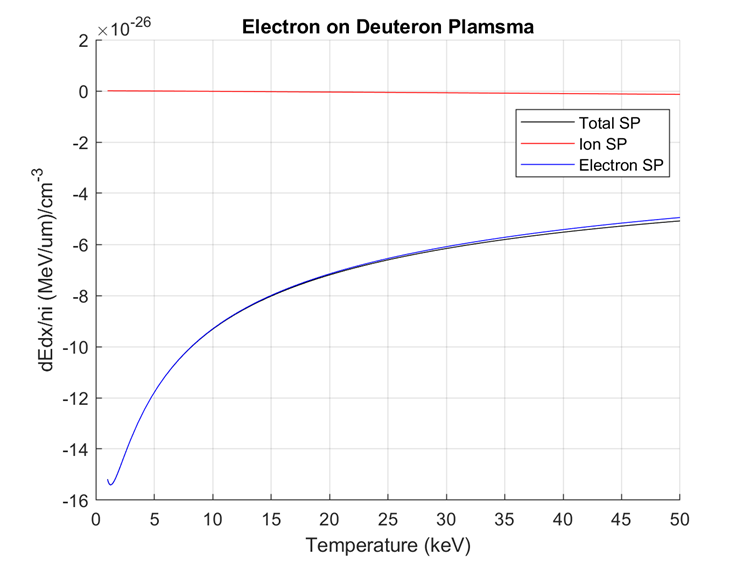}
\caption{}
\label{e-d}
\end{subfigure}
\caption{
(a) The stopping power of ${}^{11} B$ on the $p$ plasma. The projectile energy is 7.5 MeV so that it
is on the resonance production of the $ {}^{11}B + p \rightarrow 3 \alpha$ reaction. (b) The stopping power of electrons on the $t$ plasma where the electron velocity is the same as the $d$ in Fig.~\ref{fig:subim1}.
(c) The stopping of electrons on the ${}^{3}H_e$ plasma where the electron velocity is the same as the $d$ in Fig.~\ref{fig:subim3}. (d) The stopping power of electrons on the $d$ plasma where the electron velocity is the same as the incoming $d$ at 500 keV.}
\label{e-Ion}
\end{figure}

For completeness, we also plot the stopping power for the ${}^{11} B$ on $p$ plasma at the resonance
in Fig.~\ref{fig:B11_p}. In this case, the incoming energy of ${}^{11} B$ is at 7.5 MeV. We see that the 
stopping power is about an order of magnitude larger than that of $p$ on  ${}^{11} B$ plasma at $T = 10$ keV. \\

To maintain neutrality of the plasma, one will wish to accompany the ion beam with the electron beam at the same charge flux with the same charge density and velocity. In view of this, we calculate the stopping power for the $ e$ on $ t$ and
$e$ on ${}^3H_e $ with velocities that correspond to the incoming $d$ on the resonances of the $d + t$ and $d + {}^3H_e$ reactons. For the case of $e$ on $d$, the electron velocity is the same as the incoming $d$ at 500 keV. They are plotted in Fig.~\ref{e-T} for the $e$ on the $t$ plasma, Fig.~\ref{e-He3} for the $e$ on the ${}^3H_e $ plasma, and Fig.~\ref{e-d} for the $e$ on the $d$ plasma. We see that the stopping powers are negative which means the electrons gain energies in all these cases. This is because the incoming electron velocities are at 1.07\%, 2.16\% and 2.3\% of the speed of light for the three cases considered here. They are lower than the average velocity of the electrons at 7.7\% c in the plasma at as low a temperature as 1 keV. Thus, the accompanying electron beam gains energy as it traverses the plasm. Furthermore, the absolute values of the stopping powers of the electrons are larger than those of the incoming ions so that the electron beams  will be thermalized to the plasm temperature with $ T> 1$ keV faster than the thermalization of the ions beams. \\

Finally, we visit the original idea of ABRF which relies on the criterion that $R_{sp}$ in Eq.~(\ref{R-sp}) is greater than unity at certain temperature. Based on our present calculation, we find that the combined electron and ion contributions lead to $R_{sp}  = 0.42$ for $d$ on the $t$ plasma at $T = 10$ keV which is less than unity. This is smaller than that predicted in Ref.~\cite{Liu:2017pww}, which is partly due to the fact that the present calculation shows that the stopping power dependence on the plasma temperature is closer to $T^{-1/2}$ or $T^{-1}$ which does not fall as fast as the presumed $T^{-3/2}$ behavior in Ref.~\cite{Liu:2017pww}. Furthermore, the plasma ion contribution to the stopping power is not negligible as assumed before~\cite{Liu:2017pww}. Given the presently calculated $T$ dependence in Fig.~\ref{fig:subim2}, $R_{sp}$ will reach unity at $T \sim 60$ keV. For 
$d$ on ${}^3H_e $ plasma, $R_{sp} = 0.12/0.21$ at $T = 10/30$ keV. For $p$ on ${}^{11}B$ plasma, $R_{sp} = 0.063$ at $T = 30$ keV. From these calculations, it turns out that the original proposal
which relies on the idea that fusion power generation can be greater than the stopping power loss of the beam at certain reasonable temperature (e.g., less than 30 keV) does not pan out. However, the stopping power is not totally lost through plasma radiation, it can heat up the plasma or maintain its temperature. The ABFR framework can still work If part of the energy due to the stopping power loss can be tapped.
A scheme where the beam is utilized to prolong the lifetime of  the plasma and maintain the plasm temperature in a magnetic mirror so that the energy can be generated from both the resonance and the thermalnuclear fusion productions is being investigated~\cite{Liu2023}.

\section{Conclusions}

We have calculated the stopping powers of the ion and electron beams on plasmas 
based on the  analytic formulation of Brown, Preston and Singleton~\cite{Brown:2005ji} to study the temperature dependence of the $d$ on $t$ and $ {}^3 H_e$ plasmas and $p$ on ${}^{11}B$ plasma
on the resonances of their respective fusion reactions. We also studied $d$ on $d$ plasma at
500 keV incoming $d$ energy. We found that the temperature dependence of the stopping powers do not
fall as fast as $T^{-3/2}$ when the quantum effects are included. In fact,  it decreases slower, such as $T^{-x}$ with $x \le 1$ in the range of T from $\sim$ 5 to 50 keV.  

As a result of this temperature dependence and the inclusion of the non-negligible contributions  of the ions in the plasma, the ratio $R_{sp}$ in Eq.~(\ref{R-sp}) are less than unity for the cases we have considered. In other words, the fusion power generation is not enough to overcome the power loss due to the stopping power. This renders the original ABRF proposal not feasible. However, the stopping power is not totally lost. It will heat up the plasma. One will need to consider tapping this part of the energy and other mechanisms to include both the resonance and thermalnuclear fusion reactions for the ABFR to work.

\section{Acknowledgments}
We thank A. Sefkow, J. Santarius, Z.M. Sheng, and C. Crawford  for useful discussions.

\end{document}